%% file: ms_parametersv2.tex
\documentclass[apj]{emulateapj}

\usepackage{amsmath}
\usepackage{natbib}
\usepackage{graphicx}
\usepackage{epsf}
\usepackage{color}
\usepackage{threeparttable}
\usepackage{comment}
\usepackage{epsfig}
\usepackage{xspace}
\usepackage{subfigure}
\usepackage{ulem} 
\usepackage[usenames,dvipsnames,svgnames]{xcolor}
\usepackage{tabularx,ragged2e}
\newcolumntype{x}{>{\Centering}X}
\DeclareGraphicsExtensions{.jpg,.pdf,.png,.eps,.ps}

\newcommand{\PTEPlanckfssptPcnt}{\ensuremath{3.2\%}\xspace}
\newcommand{\PTEPlanckfsspt}{\ensuremath{0.032}\xspace}
\newcommand{\PTEPlanckfssptmid}{\ensuremath{0.094}\xspace}
\newcommand{\PTEPlanckfssptmin}{\ensuremath{0.24}\xspace}

\newcommand{\PTEPlanckfsplanck}{\ensuremath{0.31}\xspace}
\newcommand{\PTEPlanckfsplanckmin}{\ensuremath{0.29}\xspace}

\newcommand{\PTEPlanckfscross}{\ensuremath{0.18}\xspace}
\newcommand{\PTEPlanckfscrossmin}{\ensuremath{0.19}\xspace}

\newcommand{\PTEsptcross}{\ensuremath{0.57}\xspace}
\newcommand{\PTEsptcrossmid}{\ensuremath{0.66}\xspace}
\newcommand{\PTEsptcrossmin}{\ensuremath{0.74}\xspace}

\newcommand{\PTEsptplanck}{\ensuremath{0.20}\xspace}
\newcommand{\PTEsptplanckmid}{\ensuremath{0.38}\xspace}
\newcommand{\PTEsptplanckmin}{\ensuremath{0.32}\xspace}

\newcommand{\PTEcrossplanck}{\ensuremath{0.73}\xspace}
\newcommand{\PTEcrossplanckmin}{\ensuremath{0.62}\xspace}

 \newcommand{\LCDM}{\mbox{$\Lambda$CDM}\xspace}

 \renewcommand{\vec}[1]{\mathbf{#1}}
 \newcommand{\Lk}{\mathcal{L}}

 \newcommand{\planck}{\textit{Planck}}

 \def\sqdeg{\ensuremath{{\rm deg^2}}}

\def\be{\begin{equation}}
\def\ee{\end{equation}}
\def\ba{\begin{eqnarray}}
\def\ea{\end{eqnarray}}

\begin{document}
\title{A Comparison of Cosmological Parameters Determined from CMB Temperature Power Spectra from the South Pole Telescope and the \planck\ Satellite}

\input spt_authorlist_v2.tex

\begin{abstract}

The \planck\ cosmic microwave background (CMB) temperature data
are best fit with a \LCDM model that is in mild tension with constraints from other cosmological probes. The South Pole Telescope (SPT) 2540 $\text{deg}^2$ SPT-SZ survey offers measurements on
sub-degree angular scales (multipoles $650 \leq \ell \leq 2500$) with sufficient precision to use as an independent check of the \planck\ data. Here we build on the recent joint analysis of the SPT-SZ and \planck\ data in \citet{hou17} by comparing \LCDM parameter estimates using the temperature power spectrum from both data sets in the SPT-SZ survey region. We also restrict the multipole range used in parameter fitting to focus on modes measured well by both SPT and \planck, thereby greatly reducing sample variance as a driver of parameter differences and creating a stringent test for systematic errors.
We find no evidence of systematic errors from such tests.
When we expand the maximum multipole of SPT data used, we see low-significance shifts in the angular scale of the sound horizon and the physical baryon and cold dark matter
densities, with a resulting trend to higher Hubble constant.
When we compare SPT and \planck\ data on the SPT-SZ sky patch to \planck\ full-sky data but keep the multipole range restricted,
we find differences in the parameters $n_s$  and $A_se^{-2\tau}$.
We perform further checks, investigating instrumental effects and modeling assumptions, and we find no evidence that the effects investigated are responsible for any of the parameter shifts.
Taken together, these tests reveal no evidence for systematic errors in SPT or \planck\ data in the overlapping sky coverage and multipole range and, at most, weak evidence for a breakdown of \LCDM or systematic errors influencing either the \planck\ data outside the SPT-SZ survey area or the SPT data at $\ell >2000$.
\end{abstract}


\maketitle

\section{Introduction}

\setcounter{footnote}{0}

Anisotropies in the cosmic microwave background (CMB) have provided a wealth of information about the universe.
The CMB temperature anisotropy power spectrum in particular provides some of the tightest current constraints
on cosmological models.
The most precise measurement of the CMB temperature power spectrum at medium and large angular scales has been made by the \planck\ satellite as published in the February 2015 \planck\ data release \citep{planck15-13}.
Sensitive measurements of the CMB temperature anisotropy have also been made using ground-based telescopes such as the South Pole Telescope \citep[SPT,][]{carlstrom11} and the Atacama Cosmology Telescope \citep[ACT,][]{swetz11}.
\citet[hereafter S13]{story13} used SPT data from the 2540 deg$^2$ SPT-SZ survey to make the most precise measurement of the CMB temperature power spectrum damping tail above angular multipoles $\ell \sim 2000$ and a measurement at $650 \leq \ell \leq 2000$ second only to \planck\ in precision.

With the exquisite precision of the \planck\ measurements, signs of moderate tension have been noted between cosmological parameters estimated from the \planck\ CMB power spectra and other cosmological measurements.
For example, in \cite{riess16}, the value of $H_0$ determined from measurements of type Ia supernovae, calibrated with Cepheids, is found to be inconsistent with the \planck\ value by 3$\sigma$.
Additionally, the amplitude of density fluctuations in the local universe implied by \planck\ CMB power spectrum data
disagrees with certain local measurements of the density fluctuations at the $2 \sigma$ level
\citep[e.g.,][]{kilbinger13,planck15-24}.

Some tension has also been noted between the cosmological parameter constraints from
measurements of the CMB using different instruments or multipole ranges.
Many authors, including \citet{calabrese17}, have demonstrated 1-2$\sigma$ differences
in the values of $H_0$ and $\sigma_8$ between pre-\planck\ and \planck\ data.
\cite{addison16} point out discrepancies between cosmological parameters determined from two halves
of the \planck\ data (split at $\ell = 1000$), and between the best-fit cosmologies of \planck\ and SPT,
although \cite{planck16-51} argue that these discrepancies are statistically insignificant.

While the statistical significance of these reported discrepancies ranges from low to moderate, they could be hints of the \LCDM model breaking down or systematic contamination in one or more of the measurements.
Because \planck\ and SPT provide the most precise temperature power spectrum measurements, it is particularly important to carefully investigate any differences between these two data sets.

In a previous paper, \citet[][hereafter H17]{hou17} compared the SPT-SZ and \planck\ data at the map
and power-spectrum level, in a study similar to that performed by \citet{louis14} on ACT and \planck\
data. H17 used the \planck\ 143~GHz and SPT 150~GHz maps to create three sets of binned power spectrum measurements, or ``bandpowers,'' in the SPT-SZ sky patch, namely: the cross-spectrum of two independent halves of the \planck\ 143~GHz data ($143\times143$), the cross-spectrum of the SPT 150~GHz data and \planck\ 143~GHz data ($150\times143$), and the cross-spectrum of two independent halves of the SPT 150~GHz data ($150\times 150$).
We will refer to these collectively as the ``in-patch'' bandpowers.
In H17, these bandpowers were shown to be consistent with each other and marginally consistent with the bandpowers obtained from the full \planck\ map.
In this paper, we extend this comparison to the cosmological parameters obtained from these bandpowers
and the \planck\ full-sky power spectrum.

We start by comparing the best-fit parameters obtained from the full-sky \planck\ data
and the best-fit parameters from the SPT-SZ data, with a null hypothesis that the \LCDM model is correct and
the statistical models of both data sets are accurate.
Under this null hypothesis, the parameters derived from the two datasets are marginally discrepant---the $\chi^2$ values for these parameter differences should be larger only \PTEPlanckfssptPcnt of the time (the probability to exceed the $\chi^2$ between the parameter sets is \PTEPlanckfsspt; see Section 3 for details).
Assuming the null hypothesis is correct, this \PTEPlanckfssptPcnt probability must be understood as resulting from a somewhat (but not highly) unlikely statistical fluctuation.
Other possible explanations include uncharacterized systematic errors or a breakdown in \LCDM.
In this paper, we attempt to discriminate between these three possibilities.

After quantifying the discrepancy in parameters determined from the full SPT and \planck\ temperature power spectra,
we test for systematics by restricting the SPT-SZ and \planck\ datasets to the anisotropy modes measured well in both datasets. Specifically, we restrict both datasets to the SPT-SZ footprint, using the H17 in-patch
bandpowers, and consider a fixed multipole range. Such a restriction greatly reduces the covariance of parameter differences given our null hypothesis by eliminating nearly all the sample variance contribution.

After testing for systematic errors with the restricted data sets, we test for other potential sources of the observed parameter differences between the full data sets.
We first explore how parameters shift when the in-patch bandpowers are restricted to different ranges of angular scales; this tests for scale-dependent systematics or an inadequate cosmological model.
Next we study how parameters shift from the in-patch bandpowers to the full-sky \planck\ bandpowers; this tests if the SPT-SZ patch is sufficiently unusual to challenge the assumptions of statistical isotropy and Gaussianity underlying our cosmological model or if there are systematic errors in the \planck{} data outside of the SPT-SZ patch.
We also discuss influence on parameters of the tilt in the in-patch bandpowers relative to \planck\ full-sky data first noted in H17.

Finally, we explore several other factors that could cause the mild tension between the \planck- and SPT-derived parameters.
We examine the SPT foreground model, the SPT calculation of beam uncertainty, the SPT $\tau$ prior, and the effects of lensing.
These tests probe analysis and uncertainty modeling choices that could introduce systematic differences.

This paper is organized as follows. In Section 2 we describe our methodology for parameter estimation and comparison. In Section 3 we explore the consistency of \planck\ and SPT parameters estimated from the full data sets and from data restricted to the same sky patch and multipole range. In Section 4 we test for sources of systematic error from the foregrounds and beam uncertainty. We also discuss the influence of lensing and the $\tau$ prior on the parameter estimates. The conclusions are presented in Section 5.

\section{Parameter Estimation and Comparison Methodology}

\subsection{Bandpowers}
Several of our tests in this work make use of the publicly available \planck\ 2015 baseline high-$\ell$ temperature and low-$\ell$ temperature and polarization bandpowers.  These are optimally combined multi-frequency bandpowers, and we refer to these as the \planck{} Full Sky (PlanckFS) bandpowers \citep{planck15-11}. The \planck\ parameters we use are obtained from the baseline \LCDM Monte Carlo Markov Chain (MCMC, \citealt{planck15-13}). We also use the designation PlanckFS to refer to the parameter estimates from this chain.

We also use bandpowers created from the SPT-SZ 150~GHz and \planck\ 143~GHz maps of the 2540\,\sqdeg{} SPT-SZ survey region.
 We refer to the cross-correlation of two half-depth maps as either $150\times150$ or $143\times143$.
 The cross-spectrum $150\times143$ is the correlation of the full-depth SPT and full-depth \planck\ maps.
 A detailed description of the creation of these bandpowers is provided by H17.

\subsection{Cosmological Parameter Likelihood}

\begin{figure*}[!tbp]
\begin{centering}
\includegraphics[width = \textwidth]
{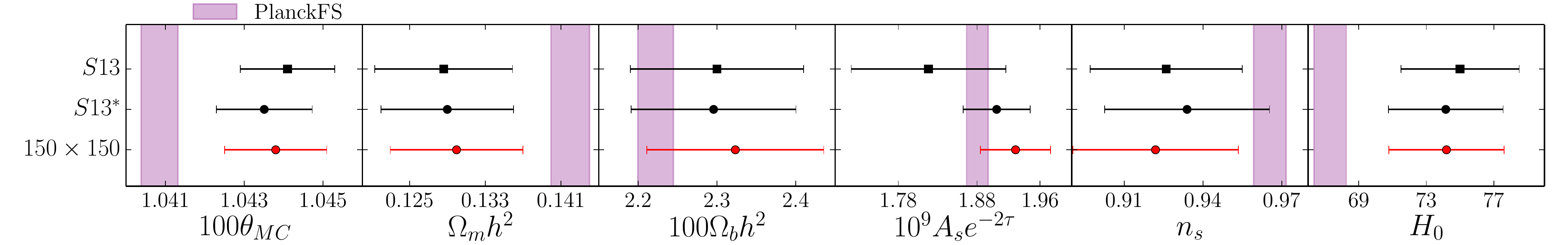}
\caption{The parameter estimates for S13, S13* (obtained from the same bandpowers as S13 but with the likelihood modifications discussed in Section \ref{sec:likechange}), and $150\times 150$ (H17). The vertical bars are the $1\sigma$ PlanckFS parameter constraints.
The estimates are based on the multipole range of $650\leq \ell \leq3000$. The shift in $A_se^{-2\tau}$ and the reduction in the error bar on that parameter combination, between S13 and S13*, come from a combination of the new calibration constraint from H17 and the correction of a bias in the calibration uncertainty treatment. The shift in $n_s$ comes from the correction of the beam uncertainty bias. The shift in $\theta_{MC}$ is primarily due to the inclusion of aberration effects.}
\label{fig:spt_1d_params}
\end{centering}
\end{figure*}

We obtain parameter estimates  by searching the space defined by the likelihood of the cosmological and nuisance parameters $\vec{\Theta}$ given the data $\vec{D}^{i\times j}$, a set of temperature bandpowers where $i$ and $j$ both run over the two frequency bands (143~GHz for \planck\ and 150~GHz for SPT). We assume the likelihood to take the following form
\begin{equation}
\begin{split}
-2\ln\Lk(\vec{\Theta}|\vec{D}^{i\times j}) = (D^{i\times j}_b-M^{i\times j}_b)(\Sigma_{bb'}^{i\times j})^{-1}(D^{i\times j}_{b'}\\-M^{i\times j}_{b'}),
    \label{eqn:likelihood}
\end{split}
\end{equation}
\noindent where  $\Sigma_{bb'}^{i\times j}$ is the bandpower covariance, and the model temperature bandpowers are expressed as
\begin{equation}
     M^{i\times j}_b = Y^iY^jW_{b\ell}^{i\times j}(D^{th}_{\ell} + F_{\ell}^{i\times j})a_{\ell} = W_{b\ell}^{i\times j}M^{i\times j}_{\ell}
     \label{eqn:model}
\end{equation}
\noindent for $i,j \in [143,150]$, and we have ignored the normalization constant in the likelihood. The $Y^{i},Y^{j}$ terms are temperature calibration parameters , $W_{b\ell}^{i\times j}$ is the bandpower window function \citep[e.g.,][]{knox99}, and $F_{\ell}^{i\times j}$ is the foreground model from \cite{story13} with frequency dependence included  \citep{george15}. The term $a_{\ell}$ includes aberration effects  due to our proper motion with respect to the CMB as
\begin{equation}
a_{\ell} = 1- \frac{d\ln C_{\ell}}{d\ln \ell}\beta\left<\cos\theta\right>,
\end{equation}
based on \cite{jeong14} and with $\beta=1.23 \times 10^{-3}$  and $\left<\cos\theta\right>=-0.26$.

The calibration parameters are based on work from H17 where the three in-patch bandpowers are simultaneously calibrated to each other over the common multipole range. The calibration parameters are fixed to 1 for $Y^{143}$ and have a Gaussian prior of $Y^{150}= 0.9914\pm0.0017$.\footnote{Note that our comparisons between parameters from 150x143 and 143x143 with parameters from PlanckFS have inconsistent treatments of 143~GHz calibration uncertainty.
In the former two the uncertainty is set to zero, while in the latter the 0.07\% absolute calibration uncertainty reported by the Planck team is included.
We expect that these inconsistent treatments have negligible impact on our results due to the fact that a 0.07\% map-level calibration uncertainty would be a highly sub-dominant contribution to any of our in-patch parameter uncertainties; for example, the fractional uncertainty on $A_{\rm s}e^{-2\tau}$ is approximately 3\% for 143x143.
We also find that fixing the relative calibration uncertainty between Planck and SPT has negligible impact on our parameter comparisons.}

We use the May 2016 version of CAMB \citep{lewis02b} to calculate the theoretical prediction for the lensed temperature power spectrum ($D^{th}_{\ell}$) assuming the standard \LCDM model. Our free parameters are: the  approximation to the
angular scale of the sound horizon, $\theta_{MC}$, the baryon and total matter densities, $\Omega_bh^2$ and $\Omega_mh^2$, the scalar amplitude, $A_se^{-2\tau}$,
and the scalar spectral index, $n_s$.
While the Hubble constant $H_0$ is derived from these five parameters, we discuss it throughout this work
because the discrepancy between the CMB-determined value of $H_0$ and local measurements is of particular interest.
We place a Gaussian prior on the optical depth $\tau$ of $0.07\pm0.02$.

Our parameter vector also includes six nuisance parameters.
The $Y^iY^j$ term is treated as a single parameter, and we include five parameters for foregrounds (three for the template amplitudes and two for the frequency dependence), giving $\vec{\Theta}$ a total of twelve elements.
All other parameters are fixed to the baseline model values from \cite{planck15-13} which are the default settings for the May 2016 version of CAMB.

\subsection{The Covariance for the In-patch Bandpowers}
The covariances for the bandpowers $150\times150$, $143\times143$, and $150\times143$ contain sample and noise variance along with beam uncertainty. A correlation matrix is formed for each source of beam uncertainty as
\begin{equation}
    \rho^{B, i\times j}_{\ell\ell'} = \bigg(\frac{\delta D_{\ell}}{D_{\ell}}^{B,i\times j}\bigg)\bigg(\frac{\delta D_{\ell'}}{D_{\ell'}}^{B,i\times j}\bigg)
    \label{eqn:beam_corr}
\end{equation}
\noindent where
\begin{equation}
     \frac{\delta D_{\ell}}{D_{\ell}}^{B,i\times j} = \bigg(1+\frac{\delta B_{\ell}}{B_{\ell}}^i\bigg)^{-1}
\bigg(1+\frac{\delta B_{\ell}}{B_{\ell}}^j\bigg)^{-1} - 1
\label{eqn:beam_temp}
\end{equation}
\noindent and $\frac{\delta B_{\ell}}{B_{\ell}}^i$ is either the \planck\ 143~GHz or the SPT-SZ 150~GHz fractional beam uncertainty template. The SPT beam functions ($B_{\ell}$) and their uncertainty ($\delta B_{\ell}$) are determined using a combination of maps from Jupiter, Venus, and
the 18 brightest point sources in the SPT patch.
We refer the reader to S13 and \cite{schaffer11} for more details on the creation of the beam templates. The correlation matrix $\rho^{B, i\times j}_{\ell\ell'}$ is then added into the full covariance as
\begin{align}
    \Sigma_{bb'}^{i\times j} &= \Sigma_{bb'}^{S,i\times j}+\Sigma_{bb'}^{N,i\times j}+\Sigma_{bb'}^{B,i\times j} \\
    \Sigma^{B,i\times j}_{bb'} &= W_{b\ell}^{i\times j}W_{b'\ell'}^{i\times j}M^{i\times j}_{\ell}M^{i\times j}_{\ell'}\rho^{B, i\times j}_{\ell\ell'},
     \label{eqn:beam_cov}
\end{align}
\noindent where $S$, $N$ and $B$ superscripts signify the sample, noise, and beam covariances respectively.

\cite{planck15-11} show that the uncertainty in the \planck\ beams has less than a 0.2\% impact on $143\times143$ bandpowers, and the resulting impact on parameter estimation is also extremely small
\citep{planck15-13}.
We thus make the simplifying assumption that $\frac{\delta B_{\ell}}{B_{\ell}}^{143} = 0$.

\subsection{Changes to Likelihood Since S13}
\label{sec:likechange}

In 2013, S13 presented  the SPT-SZ 150~GHz bandpowers and resulting \LCDM parameter constraints.
There are some differences between the parameter estimation for S13 and for this analysis. First, in this work we assume massive neutrinos with a total mass of 0.06~eV and a \planck-based $\tau$ prior. The work in \cite{calabrese17} points out the importance of these assumptions when comparing parameters estimated from the \planck\ data with other CMB results.

We handle calibration uncertainty by accounting for it in our model instead of including it in the bandpower covariance, and we use a different calibration prior than used in S13. The method outlined above for handling beam uncertainty differs from S13 in that the beam covariance is now formed in a model-dependent way based on $M_{\ell}^{i\times j}$,
the cosmology of each MCMC sample. The previous methods for handling calibration and beam uncertainty used in S13 produced biased parameter constraints, lowering $A_se^{-2\tau}$ and $n_s$ (see the Appendix for further detail).
The new, much tighter calibration prior is based on the in-patch bandpower comparisons in H17.

In S13, aberration effects due to our proper motion with respect to the CMB were not included in parameter estimation. Based on \cite{jeong14} we include aberration in Eq. 2. Accounting for aberration leads to an approximately 0.3$\sigma$ shift in the S13 $\Theta_{MC}$  value towards the PlanckFS value.

In Figure \ref{fig:spt_1d_params}, we show the differences in parameter estimates due to the above changes by comparing the parameters from S13 to parameters estimated from the same bandpowers but with the updated likelihood (S13*). The decrease in $\theta_{MC}$ is primarily the result of including aberration effects, $A_se^{-2\tau}$ is increased by the new calibration prior, and $n_s$ is mostly increased by the new method of handling beam uncertainties. The changes to the likelihood relative to S13 lead to greater consistency between \planck\ and SPT.

In H17 and this work, we use the $150 \times 150$ bandpowers generated from half-power SPT maps, instead of the bandpowers from S13, which were generated from cross-spectra of hundreds of single-observation maps.  This choice makes the data easier to simulate, and simplifies the 143 GHz cross spectrum analysis, since that data was created in a similar manner. There was no significant difference found between the $150\times150$ and S13 bandpowers---for more details see the Appendix of H17. In Figure~\ref{fig:spt_1d_params}, we compare the differences between parameters estimated from S13, S13* (the S13 bandpowers with the updated likelihood) and $150\times150$ (H17).

\subsection{Parameter Comparison and Parameter-difference Covariance}
\label{sec:parameter_comparison}

To obtain parameter estimates for our in-patch bandpowers we use the Metropolis-Hastings algorithm to produce a chain from which we generate the posterior for $\vec{\Theta}$ and then marginalize over the nuisance parameters (foreground and calibration parameters). We generate the chains using the likelihood sampler Cosmoslik \citep{2017ascl.soft01004M}.

The primary statistic we use to infer the compatibility between various parameter distributions is
\begin{equation}
 \chi^2 = \Delta \vec{\theta}^T\vec{C}^{-1}\Delta \vec{\theta}
 \label{FS_chi2}
\end{equation}
where $\vec{C}$ is the parameter difference covariance and
\begin{equation}
\Delta \vec{\theta} = \vec{p}_1 - \vec{p}_2.
\end{equation}
The $\vec{p}_{\alpha}$ are either the means of the parameter posteriors or obtained through minimization of the negative log-likelihood. The latter method is used
when simulations are required as minimizing the negative log-likelihood requires significantly less computation time than running an MCMC. The $\vec{p}_{\alpha}$ are composed of  the five non-$\tau$ cosmological parameters: $\theta_{MC}$, $\Omega_mh^2$, $\Omega_bh^2$, $A_se^{-2\tau}$, and $n_s$.

When comparing parameters from the in-patch bandpowers to PlanckFS the parameter difference
covariance is approximated as
 $\vec{C} = \vec{C}^{i\times j,\ell_{\rm{max}}} + \vec{C}^{FS}$ with $i\times j \in \{143\times 143, 150\times 143, 150\times 150\}$ and $2000\leq\ell_{\rm{max}}\leq3000$. The small
correlations between the in-patch and full-sky parameter sets are ignored.

The parameter difference covariance for comparisons between the in-patch bandpowers cannot be calculated as simply. The parameters are obtained from bandpowers in the same sky cut, and therefore a large portion of the sample variance is common between all three sets and must be accounted for. It is necessary to estimate the covariance from the fluctuations across a set of simulations.

To calculate the in-patch parameter difference covariance matrices we generate 400 bandpower simulations for each of the in-patch spectra. The creation of the simulations is described in H17. To simulate the calibration uncertainty we multiply each simulation by a random draw from the appropriate calibration prior. The simulated bandpowers are then substituted into Eq. \ref{eqn:likelihood} and we calculate a set of parameter estimates through minimization of the negative log-likelihood. The minimization is done using the \textit{scipy}
minimize module with the Nelder-Mead method \citep{scipy}. Running the minimizer on the $150\times150$ bandpowers returns parameter values similar to the results obtained from the MCMC procedure.

With the 400 sets
of parameters for each of the three in-patch bandpowers we obtain the parameter difference covariances
for the three in-patch comparisons. The stability of our
covariances was tested by splitting the simulations into two groups of 200 and recalculating $\chi^2$
with each half. We find the results from the two halves to be consistent, and, as we show below, the simulated parameter differences follow a $\chi^2$ distribution with five degrees of freedom.
After calculating the $\chi^2$ for a set of parameter differences we convert it to a probability to exceed (PTE) which we use to infer compatibility between the sets of parameters.

\section{SPT-SZ and \planck\ Consistency Tests}
\label{sec:consistency_tests}
In this section,
we use the methodology of Section \ref{sec:parameter_comparison} to quantify
the significance of differences in parameters estimated from SPT-SZ data and \planck\ data.
Our primary metric is a $\chi^2$ statistic and its associated PTE.
We first compare the parameter constraints from the PlanckFS and SPT $150\times150$ data sets
and find a relatively low PTE of \PTEPlanckfssptPcnt. We then perform a series of tests investigating
possible causes for this low PTE.

In Section~\ref{sec:in_patch}, we test the hypothesis of a systematic error in one or both experiments by restricting the \planck\ and SPT datasets to modes on the sky that are measured well by both experiments. In particular, we restrict the \planck\ data to the SPT-SZ patch, and only consider multipoles in the range $650 \leq \ell \leq 2000$. By doing so we greatly decrease the expected variance in parameter differences under our null hypothesis, primarily because we eliminate  nearly all the sample variance contribution.
The volume of parameter space within the 1$\sigma$ uncertainties in parameter differences is reduced by
a factor of over 300 relative to the comparison of the full data sets, greatly increasing sensitivity to systematic errors.

In Section~\ref{sec:multipole_range}, we re-introduce sky modes that are only measured well by one of the two experiments, either
by \planck\ outside the SPT-SZ survey region or by SPT in the multipole range above which
the in-patch \planck\ data become very noisy.
Specifically, we first explore the consistency of parameters from in-patch bandpowers over several multipole ranges by varying $\ell_{\rm{max}}$, the maximum multipole included for parameter estimation.
In Section~\ref{sec:sky_coverage}, we then compare the various in-patch bandpowers to the PlanckFS dataset, also comparing different $\ell$ ranges. We then discuss specific features of the data and parameter shifts of interest in Sections~\ref{sec:bandpower_ratios} and \ref{sec:H0}.

\subsection{SPT and \planck\ Parameter Comparison, Full Data Sets}
\label{sec:Planckfsspt}

Comparing the parameter differences derived from the SPT $150\times150$ and PlanckFS data sets, we find
\begin{equation}
  {\rm PTE} = \PTEPlanckfsspt \, (\hbox{PlanckFS vs.~$150\times150$}).
\end{equation}
 When considering parameters estimated from the full multipole range for $150\times150$, the parameters that differ the most (in terms of standard deviations) are $\theta_{MC}$ and $\Omega_mh^2$, and subsequently $H_0$.
As noted in previous publications, the larger value of $H_0$ preferred by SPT is closer to the value reported by \cite{riess16}.
While a PTE of \PTEPlanckfsspt is plausibly due to a statistical fluctuation, it could
be an indication of potential systematic error or a breakdown of \LCDM.

\subsection{Comparison of \planck\ and SPT in the SPT-SZ survey region}
\label{sec:in_patch}

\begin{figure*}[!tbp]
\begin{centering}
\includegraphics[width = \textwidth,scale = 0.6]
{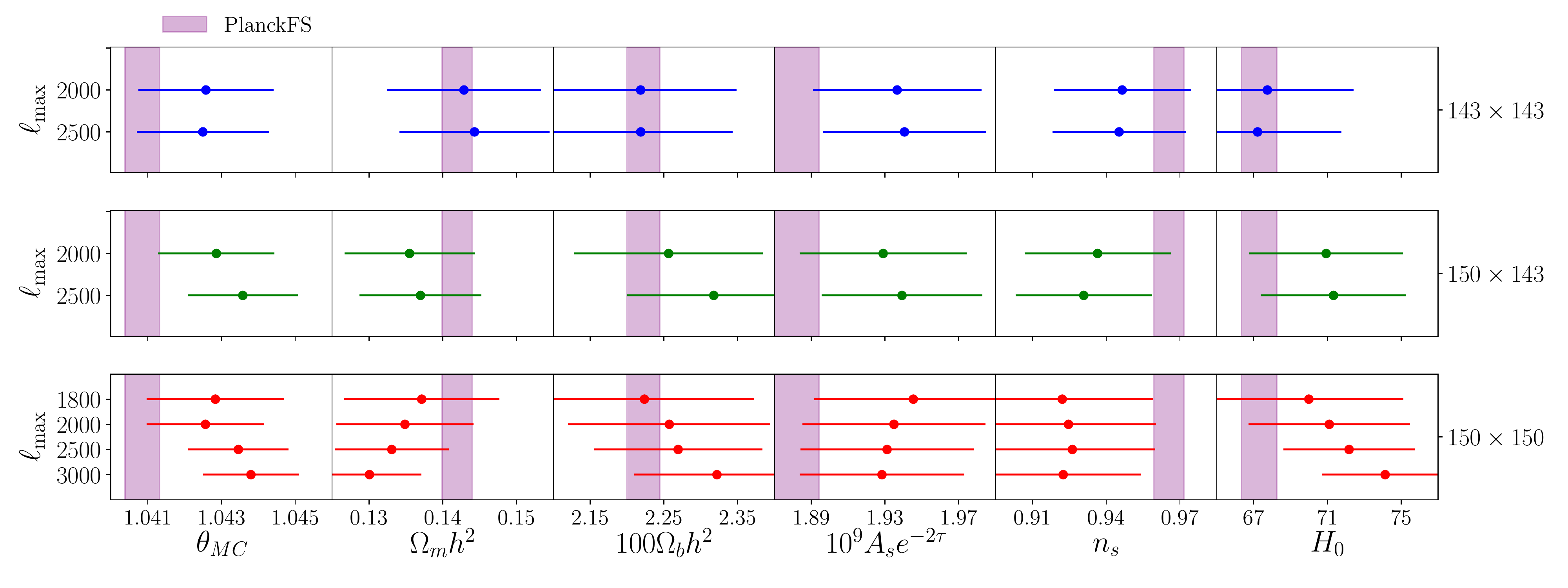}
\caption{The parameter estimates for the three sets of in-patch bandpowers for various $\ell_{\rm{max}}$  values. The estimates are based on the multipole range of $650\leq\ell\leq\ell_{\rm{max}}$. There is a noticeable trend in the $150\times150$ density parameters towards better agreement with PlanckFS as  $\ell_{\rm{max}}$ is lowered. }
\label{fig:lmax}
\end{centering}
\end{figure*}
\begin{figure*}[!tbp]
\begin{centering}
\includegraphics[width=0.96\textwidth, trim=.8cm 0 0 0]
{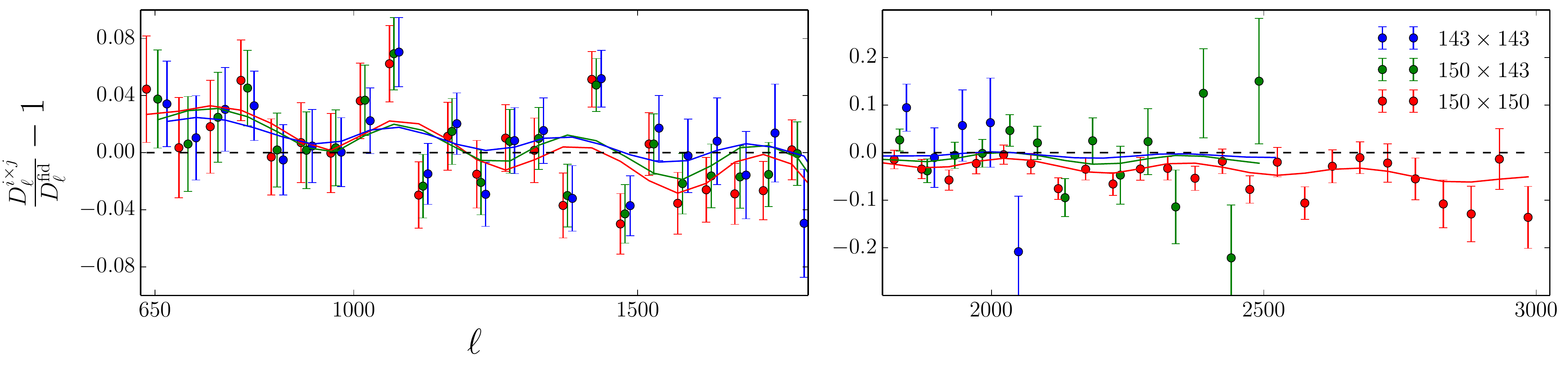}
\caption{The fractional difference between the in-patch bandpowers and a PlanckFS best-fit model. The two panels are split at $\ell=1800$ to accommodate the
growth in the $143\times143$ and $150\times143$ errorbars beyond this point. The solid curves are the best-fit models of the in-patch bandpowers divided by the best-fit PlanckFS model. Foregrounds have been subtracted. Foreground and beam uncertainty are not included in the error bars.}
\label{fig:best_fit_models}
\end{centering}
\end{figure*}

In this section, we compare parameters derived from modes that are well measured by both \planck\ and SPT.
We consider data within the SPT-SZ sky region, which covers 2540 deg$^2$, or about 6\% of the sky, and within the multipole range $650 \leq \ell \leq 2000$.
Specifically, we compare parameters estimated in this multipole range from the in-patch cross-spectrum bandpowers presented in H17.
This comparison provides a sensitive test of unaccounted-for systematic errors in either experiment.

The lower cutoff of $\ell = 650$ is set by the SPT analysis in S13, which did not report data at larger angular scales (lower multipoles) due to the increasing noise from the atmosphere on these scales.
The upper cutoff of $\ell=2000$ is set by high-$\ell$ noise in the \planck\ 143~GHz data resulting from the larger \planck\ beam (roughly 7 arcmin FWHM, compared to 1 arcmin for SPT 150~GHz) and the
slightly higher noise per pixel in the \planck\ maps ($\sim$25 $\mu$K-arcmin for \planck\ 143~GHz
compared to $\sim$18 for SPT).
The variance of the $143\times143$ bandpowers (the set with the largest noise variance)
is dominated by sample variance to approximately $\ell=1500$; as a result,
the three sets of bandpowers have similar uncertainty in this range.
The $143\times143$ errorbars begin to grow significantly larger than those for $150\times150$ around $\ell = 1800$, and $150\times143$ begins to show the same behavior around $\ell=2200$.
We choose $\ell_{\rm{max}} = 2000$ to maximize the signal-to-noise of the comparison between
$150\times150$ and $150\times143$.
When restricted to the SPT-SZ sky area and this range of angular scales, both experiments are measuring a very similar set of modes on the sky with similar signal-to-noise per mode.
Given our null hypothesis, the expected covariance of parameter \textit{differences} for these modes is thus greatly reduced, making it easier for us to see the impact of any systematic errors.

\begin{figure}[!tbp]
\begin{centering}
\includegraphics[width=0.48\textwidth]
{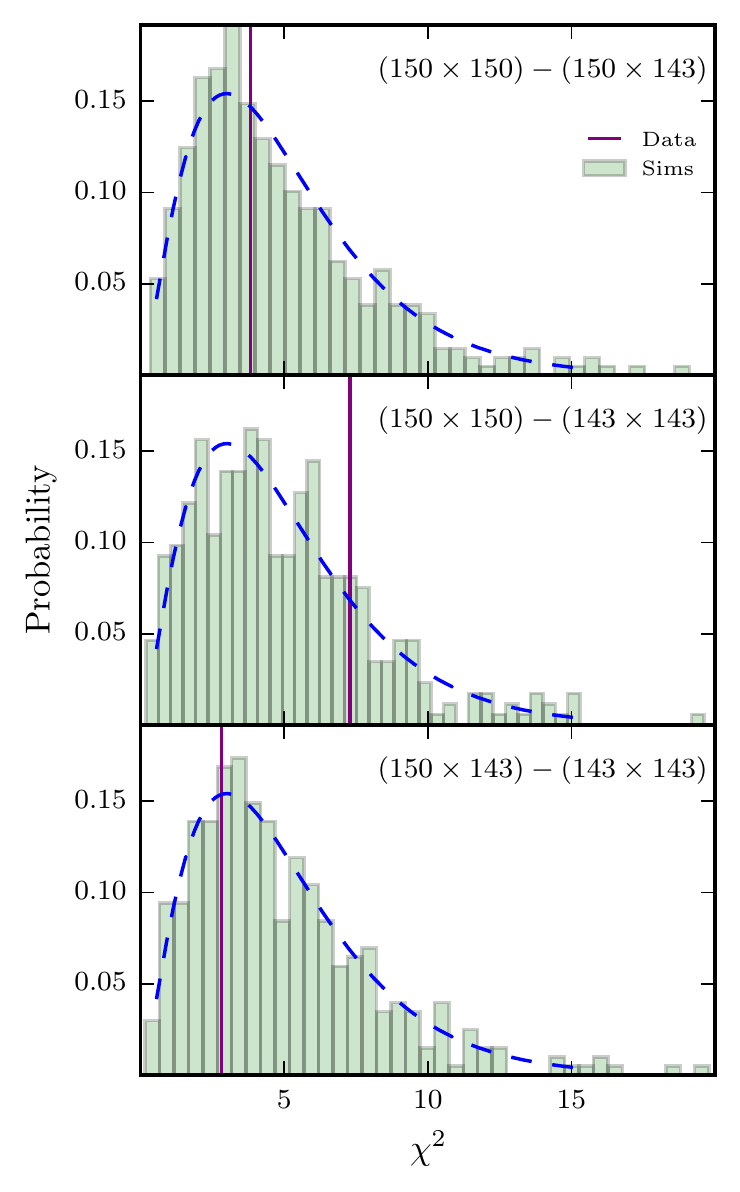}
\caption{The $\chi^2$ distributions for the three simulation differences. The vertical purple line marks the $\chi^2$ value from the data. The dashed line is a $\chi^2$ distribution for five degrees of freedom. The $\chi^2$ values are based on the multipole range of $650\leq \ell \leq2000$.}
\label{fig:sim_chi_dist}
\end{centering}
\end{figure}

The parameter estimates for the in-patch bandpowers over this multipole range are in Figure \ref{fig:lmax}, and
Figure \ref{fig:best_fit_models} shows the ratio of the in-patch bandpowers and models to the best-fit PlanckFS model.
The in-patch parameters are more similar to each other than to the \planck\ full-sky values, and the
features apparent by eye in the bandpower and best-fit-model ratios to PlanckFS are similar among
the three in-patch sets. These plots still include sample variance in the in-patch error bars, however,
so it is difficult to assess the statistical consistency of the three data sets.
Figure~\ref{fig:sim_chi_dist} shows the distribution of $\chi^2$ values for differences in simulation parameters calculated in this comparison (as expected, the histogram closely follows a $\chi^2$ distribution for five degrees of freedom).
This statistic accounts for the large decrease in sample variance in the parameter difference covariance
and provides a quantitative assessment of the consistency among the three in-patch parameter sets.
The $\chi^2$ values of the data differences are shown by vertical red lines, and none of these
values lie notably outside the main distribution.
The PTEs from this test are included in Table \ref{table:param SPT patch PTE} and confirm
that all three sets of in-patch bandpowers are consistent with each other in the multipole range
$650 \leq \ell \leq 2000.$

There are some parameter differences among the three in-patch sets visible in Figure~\ref{fig:lmax}.
Between the $143\times143$ and $150\times143$ data sets, the largest differences are the
slightly lower preferred values of $\Omega_mh^2$ and $n_s$ in $150\times143$.
This trend continues in $150\times150$, but as shown in Figure~\ref{fig:sim_chi_dist} and Table \ref{table:param SPT patch PTE},
these differences are consistent with our null hypothesis.

\begin{figure*}[!tbp]
\begin{centering}
\includegraphics[width = \textwidth]
{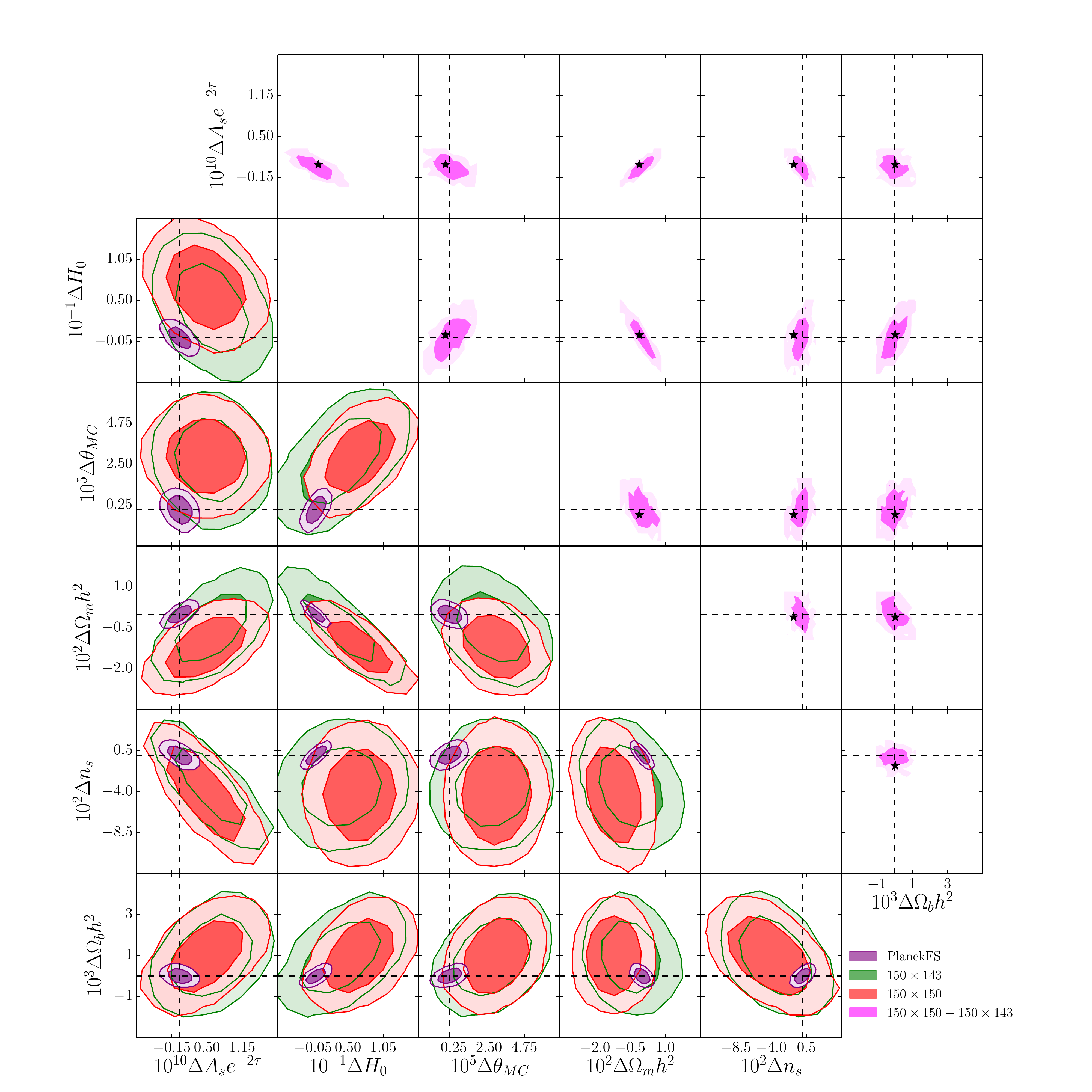}
\caption{This plot shows posterior distributions for parameter \textit{differences} for several different tests.
  In all cases, contours indicate the 68$\%$ and 95$\%$ confidence regions. The dashed lines correspond to $\Delta=0$.
  \textbf{Lower triangle:} The posterior distributions for $150\times150$, $150\times143$, both with $\ell_{\rm{max}} = 2000$,  and PlanckFS. Each distribution has the PlanckFS best-fit values subtracted.
  \textbf{Upper triangle:} Contours indicate the posterior distributions from simulations for $150\times150 - 150\times143$  with $\ell_{\rm{max}} = 2000$, and black stars indicate the parameter difference values from the same comparison in the data.
  It is visually apparent that this comparison constitutes a much more stringent consistency test than comparing to PlanckFS;
  in fact, this comparison reduces the parameter volume by a factor of $300$ (see text for details).
  The observed consistency provides strong evidence against a systematic difference in the modes measured in common between the two experiments.  }
\label{fig:dt}
\end{centering}
\end{figure*}

We pay special attention to the comparison between the parameters derived from $150\times150$ with $\ell_{\rm max}=2000$ and those from $150\times143$ with $\ell_{\rm max}=2000$, because this comparison provides the most stringent test of our null hypothesis. In Fig.~\ref{fig:dt}, one can see that the expected covariance of parameter differences, indicated by the contours in the upper triangle, are quite small, comparable to the covariance of parameter uncertainties in the PlanckFS posterior. In this regard, examining these parameter differences provides us with a much more powerful test than the comparison between the PlanckFS parameters and the  $150\times150$ full $\ell$-range parameters.

In addition to the visual impression of this increase in precision of the test given by Fig.~\ref{fig:dt}, we also provide a quantitative description of the increase in precision.
We do so by simultaneously diagonalizing the covariances for the $150\times150$ and PlanckFS parameter differences at $\ell_{\rm{max}} = 3000$ and the $150\times150$ and $150\times143$ parameter differences at $\ell_{\rm{max}} = 2000$.
We then multiply the square root of the eigenvalue ratios to calculate the reduction in the $1\sigma$ volume for the five-dimensional parameter space.
Comparing the ratio of the volumes, we find a ratio of 0.003, i.e., the volume in the parameter difference space containing 68\% of the probability is 300 times smaller for the $\ell_{\rm max}=2000$ $150\times150$ vs. $150\times143$ parameter differences than for the full $\ell$-range $150 \times 150$ vs. PlanckFS parameter differences. Despite the precision of this test, we find a perfectly acceptable PTE for this comparison:
\begin{equation}
  {\rm PTE} = \PTEsptcrossmin\, (\hbox{$150\times150$ vs.~$150\times143$}).
\end{equation}

In conclusion, when \planck\ and SPT data are restricted to modes on the sky measured well in both experiments, we find that the best-fit cosmological parameters are fully consistent.
The observed consistency between the two data sets in this stringent test provides strong evidence against instrumental systematics affecting either data set on these angular scales, on this part of the sky.

\subsection{ Relaxing Restrictions on Multipole Range}
\label{sec:multipole_range}
Next, we re-introduce smaller angular scales measured within the SPT-SZ survey region.
Given the \LCDM model and a spectrum measured in the range $650\leq \ell \leq2000$, one can predict the spectrum at other angular scales.
Given the consistency found in the previous section, finding significant tension from extending the multipole range could indicate either a systematic affecting SPT data at high $\ell$ or a failure of the \LCDM model.
Parameter estimates for the in-patch bandpowers at various values of $\ell_{\rm{max}}$ are shown in Figure \ref{fig:lmax} and PTEs are reported in Table \ref{table:param SPT patch PTE}.
 
 \begin{table}[ht]
\centering
\caption{PTEs Between Parameters in SPT Sky Patch.}
 \begin{tabular}{ c c c c }
 \hline\hline
 &&$\ell_{\rm{max}}$& \rule{0pt}{2.5ex} \\ [0.5ex]
 \hline
  & $2000$  & $2500$ &  $3000$ \rule{0pt}{3.0ex} \\ [0.5ex]
 \hline
 $150 \times 150 - 150 \times 143$ & \PTEsptcrossmin & \PTEsptcrossmid &\PTEsptcross \rule{0pt}{3.0ex} \\
 $150 \times 150-143 \times 143$ & \PTEsptplanckmin & \PTEsptplanckmid & \PTEsptplanck\\
 $150 \times 143-143 \times 143$ &   \PTEcrossplanckmin & \PTEcrossplanck & \\  [1ex]
 \hline
\end{tabular}
\label{table:param SPT patch PTE}
\end{table}

For $143\times143$, increasing $\ell_{\rm{max}}$ from 2000 to 2500 adds little information to the parameter estimates.
This is consistent with Figure \ref{fig:best_fit_models}, where the errorbars of the $143\times143$ bandpowers have become significantly larger by $\ell=1800$.
Some parameters in the $150\times143$ measurement do shift when we expand the $\ell$ range:
$\theta_{MC}$, $\Omega_bh^2$, and $n_s$ shift away from the $143\times143$ values with increasing $\ell_{\rm{max}}$.
Nevertheless, at $\ell_{\rm{max}}$ = 2500 we still find that the $150\times143$ and $143\times143$ measurements are consistent with a PTE of \PTEcrossplanckmin.\footnote{Note that for the bottom two rows of Table 1 the PTE increases as $\ell_{\rm{max}}$ is increased from 2000 to 2500. This increase is driven by the increase in the parameter difference covariances as sources of fluctuation are added that are not common to the two datasets in question---most predominantly from noise in the 143~GHz map.}

For $150\times150$, when $\ell_{\rm{max}}$ is increased from 2000 to 2500, $\theta_{MC}$ increases in a manner similar to what we saw with $150\times143$, the baryon density increases, and the matter density decreases.
These shifts correspond to an increase in $H_0$.
At $\ell_{\rm{max}} = 2500$, the $150\times150$ measurement remains consistent with $150\times143$ and $143\times143$, with PTEs of \PTEsptcrossmid and \PTEsptplanckmid respectively.
The trend in parameter shifts continues when we increase $\ell_{\rm{max}}$ to 3000 to include the full range of the $150\times150$ data, yet the PTEs remain moderate. We also plot in Fig.~\ref{fig:lmax} parameter results for $\ell_{\rm max} = 1800$, and we see the trend toward the PlanckFS values continues. Uncertainties rapidly grow for $\ell_{\rm max} < 1800$. Finally, we calculate the $\chi^2$ and PTEs for the comparison of parameters from $150\times150$ data at $\ell_{\rm max} = 2000$ to parameters from $150\times150$ data with $\ell_{\rm max} = 2500$ and 3000. These PTEs are 0.88 and 0.75 respectively. Thus, while the parameter shifts with increasing $\ell_{\rm max}$ are suggestive of a potentially interesting trend, they are consistent with our expectations under the null hypothesis.

\subsection{Relaxing Restrictions on Sky Coverage}
\label{sec:sky_coverage}

In the previous sections, we have found that the \LCDM parameters estimated from the \planck{} and SPT in-patch bandpowers are consistent for all $\ell$ ranges considered.
In this section, we relax the restrictions on sky coverage and compare the in-patch parameters to the PlanckFS parameters in different fixed $\ell$ ranges.
This effectively tests \LCDM and the assumptions of statistical isotropy in the CMB.
The PTEs between the in-patch parameters and PlanckFS parameters in all $\ell$ ranges tested are listed in Table \ref{table:param PTE}
and the parameter constraints are shown in Figure~\ref{fig:lmax}.

We first compare PlanckFS to $143\times143$ at $\ell_{\rm{max}}$ = 2000 and 2500.
Because of the rapidly increasing noise at high $\ell$ in the $143\times143$ bandpowers,
we do not consider $\ell_{\rm{max}}$ = 3000 for this comparison, and we see very little difference in the parameters
and comparison PTEs for $\ell_{\rm{max}}$ = 2000 and 2500.
Those PTEs are \PTEPlanckfsplanckmin and \PTEPlanckfsplanck, respectively.
Although the PTE values indicate no tension between the data sets,
we do see small differences in $\theta_{MC}$, $A_se^{-2\tau}$, and $n_s$ between PlanckFS and $143\times143$ at $\ell_{\rm{max}}$ = 2000 and 2500.
The two main differences between the sets of bandpowers that could drive parameter differences are the \planck\ low $\ell$ data (which are not included in the in-patch bandpowers) and the sky outside of the patch.
However, we can rule out the low-$\ell$ data as an explanation based on the results of \citet{planck16-51}, who found only marginal parameter shifts when cutting the low-$\ell$ ($\ell\leq650$) data.
Therefore, the majority of the parameter differences between PlanckFS and $143\times143$ can be attributed to differences in the \planck{} data from the SPT-SZ patch and the \planck{} data from the rest of the sky.

As noted in H17, all three sets of in-patch bandpowers
have more power than the PlanckFS model at moderate $\ell$ and less power at high $\ell$, creating a tilt.
In Figure \ref{fig:best_fit_models}, we show the ratios of the in-patch bandpowers
to the PlanckFS best-fit model.
We also show the ratios of best-fit in-patch models to the PlanckFS best-fit model.
The tilt is clearly visible by eye in all bandpowers and best-fit models,
and it is this tilt that drives the differences in $A_se^{-2\tau}$ and $n_s$.
We discuss this tilt further in Section \ref{sec:bandpower_ratios}.
There is also an oscillatory
pattern in the best-fit model ratios, consistent with the difference in best-fit $\theta_{MC}$, though this
feature is not as obviously discernible directly in the bandpower ratios as is the tilt.

Next we compare PlanckFS to $150\times143$ at $\ell_{\rm{max}}$ = 2000 and 2500 (and again refer the reader to Figure~\ref{fig:lmax}).
Compared to $143\times143$,
all parameters except $A_se^{-2\tau}$ shift away from PlanckFS values when SPT data are included
and as $\ell_{\rm{max}}$ is increased.
Nevertheless, the comparison to PlanckFS still gives PTEs of \PTEPlanckfscrossmin
at $\ell_{\rm{max}}$ = 2000 and \PTEPlanckfscross at $\ell_{\rm{max}}$ = 2500.

Finally, we compare PlanckFS to $150\times150$ with varying $\ell_{\rm{max}}$.
In Fig.~\ref{fig:lmax} we see a trend toward the PlanckFS values as $\ell_{\rm max}$ is lowered, with near-convergence of $\Omega_{\rm b}h^2, \Omega_{\rm m}h^2$, and $H_0$ by $\ell_{\rm{max}} = 1800$. For $\ell_{\rm{max}}$ = 1800, 2000, and 2500, we find $150\times150$ and PlanckFS to be at least marginally consistent (minimum PTE of \PTEPlanckfssptmid).
The tension between the data sets only approaches the 2$\sigma$ level when we extend the parameter estimation to the full $150\times150$ multipole range, where we find the PTE of \PTEPlanckfsspt with which we began this investigation.
As noted in the previous section, the only notable difference between the $150\times143$ and $150\times150$ data at $\ell_{\rm{max}}$ = 2500 is the marginally lower matter density (and hence higher Hubble constant)
that $150\times150$ prefers;
these parameters are pushed even farther in this direction when $\ell_{\rm{max}}$ is increased to 3000.
From this comparison, we see that the $150\times150$ bandpowers above $\ell > 1800$ drive some of the tension with PlanckFS.

\begin{table}[ht]
\centering
\caption{PTEs Between PlanckFS and In-patch Parameters.}
 \begin{tabular}{ c c c c c }
 \hline\hline
 &&$\ell_{\rm{max}}$& \rule{0pt}{2.5ex} \\ [0.5ex]
 \hline
  & 1800 & $2000$  & $2500$ &  $3000$ \rule{0pt}{3ex} \\ [0.5ex]
 \hline
 $150 \times 150$ & 0.21 & \PTEPlanckfssptmin & \PTEPlanckfssptmid & \PTEPlanckfsspt \rule{0pt}{3ex} \\
 $150 \times 143$ & &  \PTEPlanckfscrossmin & \PTEPlanckfscross & \\
 $143 \times 143$ & & \PTEPlanckfsplanckmin & \PTEPlanckfsplanck & \\ [1ex]
 \hline
\end{tabular}
\begin{tablenotes}
      \small
      \item The entries for $150\times143$ and $143\times143$ at $\ell_{\rm{max}}=3000$ are blank since these spectra  have negligible signal-to-noise above $\ell=2500$.
      \end{tablenotes}
\label{table:param PTE}
\end{table}

\subsection{Bandpower Ratios}
\label{sec:bandpower_ratios}

\begin{figure*}
\includegraphics[width=\textwidth, trim=3cm 0 4cm 2cm]{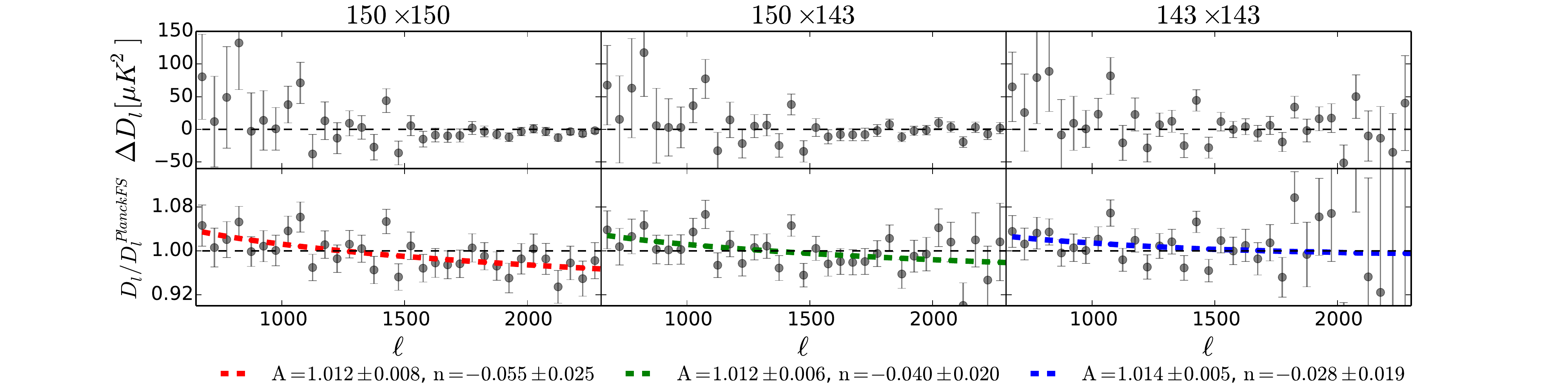}
\caption{\textit{Upper panel}: Residuals between the in-patch bandpowers and PlanckFS best-fit model. \textit{Lower panel}: Ratios of the in-patch bandpowers and the PlanckFS best-fit model. The colored dashed lines are the best-fit power law for each ratio. The uncertainty for the best-fit power-law parameters is smallest for $143\times143$ because of the small beam and calibration uncertainty for these bandpowers.}
\label{fig:bandpower_model_ratio}
\end{figure*}

In H17, it was shown that the bandpowers from the SPT-SZ patch (from both SPT and \planck\ data) have a tilt relative to the PlanckFS bandpowers.
We also see this feature prominently in the ratios of in-patch bandpowers to the PlanckFS best-fit model,
as plotted in Figure~\ref{fig:best_fit_models}.
In this section, we investigate the impact of this tilt on cosmological parameters.
To do so we fit a power law to the ratio of the in-patch bandpowers to the PlanckFS best-fit model and multiply this power law into the theory spectrum in Eq.2 when estimating new parameters.

The power law takes the form
\be
\frac{D^{th,i\times j}_{\ell}}{D^{\rm{PlanckFS}}_{\ell}}=A \left(\frac{\ell}{1000}\right)^n.
\ee
Using the PlanckFS  model instead of the PlanckFS bandpowers allows us to better assess how this tilt drives the best-fit parameters away from the PlanckFS values.
We assume Gaussianity and use the following likelihood,
\begin{equation}
-2\ln\Lk(\mathrm{A,n}|D^{i\times j}) =  \Delta_b\Sigma_{bb'}^{-1}\Delta_b'
\end{equation}

\begin{equation}
\begin{split}
\Delta_b = D^{i\times j}_b  - Y^iY^jW^{i\times j}_{b\ell}a_{\ell}\bigg(D^{\mathrm{PlanckFS}}_\ell\mathrm{A}\bigg(\frac{\ell}{1000}\bigg)^\mathrm{n}\\ + F^{i\times j}_{\ell}\bigg)
\end{split}
\end{equation}

\noindent where $D^{\mathrm{PlanckFS}}_\ell$ is the PlanckFS best-fit model. In Figure~\ref{fig:bandpower_model_ratio}, we present the best-fit power laws for each spectrum with $\ell_{\rm{max}}=2500$. As the amount of SPT data included is increased (i.e., as we go from $143\times 143$ to $150\times 143$ to $150\times 150$), we see an increase in the tilt, with $150\times 150$ having a best-fit value of $n$ that is discrepant with 0 by $2.2\sigma$. We note, however, that the best-fit tilt values from the three bandpower sets are consistent within 1$\sigma$.

To connect this feature with parameters, we remove this tilt from the $150\times 150$ bandpowers by multiplying the theory spectrum in Eq. \ref{eqn:model} by the best-fit power law for $150\times 150$ and run a new chain.
As shown in Figure \ref{fig:test_1d_params}, removing this tilt from the full range of the $150\times 150$ bandpowers results in a  decrease in $A_se^{-2\tau}$ and an increase in $n_s$ to significantly better agreement with PlanckFS.
By contrast, we note that removing the tilt hardly affects the two density parameters (and $H_0$);
therefore, the preference for higher $H_0$ by $150\times150$ is due to high-$\ell$ information unrelated to the tilt.

\subsection{Discussion of Shifts in Cosmological Parameters, and the Hubble Constant in Particular}
\label{sec:H0}

The picture that emerges from the previous sections is the following: When SPT and \planck\ data
are restricted to modes on the sky measured well in both experiments---i.e., to modes in the SPT-SZ survey
region in the multipole range $650 \leq \ell \leq 2000$---the best-fit parameters from the
two data sets are fully consistent.
When the comparison is relaxed to either just the same sky or just the same multipole range, parameter constraints from the two experiments are still marginally consistent, though parameter differences begin to emerge.
It is only when we relax all restrictions on sky coverage and multipole range that we find upwards of
2$\sigma$ tension between \planck{} and SPT.
This tension arises in roughly equal part from the SPT data at high $\ell$ and from differences in the SPT-SZ patch relative to the whole sky at moderate $\ell$.
The latter fluctuation accounts for nearly all of the difference in $A_se^{-2\tau}$, and a sizeable fraction of the differences in $\theta_{MC}$ and $n_s$, but almost none of the differences in the other parameters. The remainder of the parameter differences arise from the high-$\ell$ fluctuation.

Of the parameter differences driven by high-$\ell$ SPT data,
the Hubble constant is of particular interest given the discrepancy between the value derived from \planck\ CMB power spectrum data assuming \LCDM and the traditional distance ladder measurement of \citet{riess16}.
As can be seen in Figure \ref{fig:lmax}, half of the Hubble constant difference between SPT and \planck\ arises from the SPT data at $\ell > 2000$.

In our parameterization, the Hubble constant is a derived parameter that can be calculated from $\Omega_bh^2$, $\Omega_mh^2$, and $\theta_{\rm MC}$.
From the perspective of the Hubble constant, the angle $\theta_{\rm MC}$ is essentially fixed---the uncertainties and shifts between data sets are so small that the impact on $H_0$ is negligible.
Thus the observed variation in $H_0$ between datasets is due to changes in the two density parameters.
Changing either the baryon density or the matter density (and enforcing a flat universe) would result in a
change in the angular size of the sound horizon at recombination and, hence, a different observed value
of $\theta_{\rm MC}$. The only parameter available to preserve the observed $\theta_{\rm MC}$ is $H_0$.

Specifically, the baryon density affects the sound speed in the early universe. Increasing the baryon
density decreases the sound speed and thus the physical size of the sound horizon at recombination.
To preserve the angular size, the angular diameter distance to recombination
\begin{eqnarray}
\label{eqn:daz}
d_A(z_*) & \propto & \int_0^{z_*} dz/H(z) \\
\nonumber & \propto & \frac{1}{H_0} \int_0^{z_*} \frac{dz}{\sqrt{\Omega_{m,0} (1+z)^3 + \Omega_\Lambda}},
\end{eqnarray}
where $z_*$ is the redshift of recombination, must be made smaller. At
fixed matter density and with flatness enforced, this can only be achieved by increasing $H_0$.

Changing the matter density, meanwhile, affects the expansion rate, both in the early universe
and from recombination to today (as can be seen from Eq.~\ref{eqn:daz}).
In the early universe, this change would affect the physical size of the sound horizon,
though its impact is softened by the contribution of radiation density to the expansion rate.
Decreasing the matter density at late times would increase the
angular diameter distance to recombination, which, at fixed baryon density, would make the
angular size of the sound horizon too small. To preserve the measured angular size, $H_0$
must increase.
Thus both the increase in $\Omega_b h^2$ and the decrease in $\Omega_mh^2$ driven by the high-$\ell$ SPT data lead to an increase in the inferred value of $H_0$.

\section{Additional Tests}
\label{sec:addtests}
In the previous section, we found no evidence the parameter differences between SPT and PlanckFS at $650 \leq \ell \leq 2000$ are driven by  instrumental systematics.
In this section, we investigate potential systematic contributions to the parameter differences
driven by SPT data at $\ell > 2000$
by examining the SPT foreground model and the SPT calculation of beam uncertainty.
We also investigate other known sources of potential systematic uncertainty (not specific
to high-$\ell$ data), including the SPT $\tau$ prior and the effects of lensing.
These tests differ from those in the previous section in that they are more specific probes for systematic errors that do not aim to reduce the comparison to parameters estimated from the same modes. Instead, they focus on places where systematics may have entered into the parameter estimation, either through instrumental effects or faulty modeling assumptions.

\begin{figure*}[!tbp]
\begin{centering}
\includegraphics[width = \textwidth]
{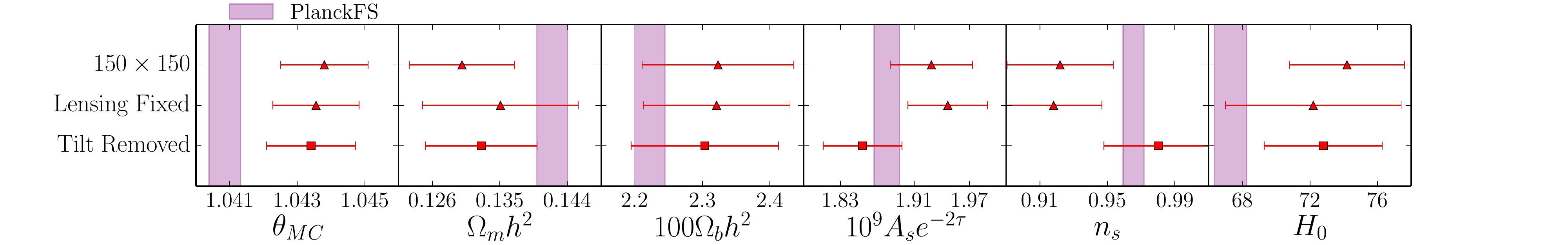}
\caption{Parameter estimates for $150\times 150$ bandpowers in the fiducial case and as a result of two tests for systematics discussed in Section~\ref{sec:addtests}.
``Lensing Fixed'': The parameter estimates from the chain with lensing fixed to the $150\times150$ best-fit. Lensing information is important for constraining the matter density. ``Tilt Removed'': The parameter estimates after removing the best-fit power law from Section 4.3 from the $150\times 150$ bandpowers. The tilt in the ratio of the $150\times150$ bandpowers to the PlanckFS best-fit model connects mostly with $A_se^{-2\tau}$ and $n_s$.
The estimates are based on the multipole range $650\leq \ell \leq3000$.}
\label{fig:test_1d_params}
\end{centering}
\end{figure*}

\subsection{Parameter Dependence on Beams and Foregrounds}

In this section, we investigate the possible impact of foreground and beam mis-estimation on parameter
differences, particularly at high $\ell$.
 To test for beam systematics, we include the amplitudes of the fractional beam uncertainty as parameters  in the MCMC, rather than analytically marginalizing over the beam uncertainty.
 What this means in practice is we modify Eq. \ref{eqn:beam_temp} to include parametrized amplitudes of the SPT beam error templates for each source of beam uncertainty and multiply this into Eq. \ref{eqn:model}. The covariance in Eq.~\ref{eqn:likelihood} no longer has Eq. \ref{eqn:beam_cov} included in it. Our model bandpowers now take the form
 
 \begin{equation}
    \label{model}
     M^{i\times j}_b = W_{b\ell}^{i\times j}M^{i\times j}_{\ell}\bigg(1+\frac{\delta D_{\ell}}{D_{\ell}}^{B,i\times j}\bigg).
\end{equation}
 
At each step of the new $150\times150$ chain we calculate the five-parameter $\chi^2$ from Eq. 8 using the difference between the cosmological parameters of the current step and the PlanckFS means. If a beam or foreground parameter were connected to the low PTE found in Section~\ref{sec:consistency_tests}, we would expect to find the $\chi^2$ posterior to be correlated with said parameter. In other words, we treat the five-parameter $\chi^2$ as a derived parameter and
look for correlations or degeneracies between this derived parameter and the beam or foreground parameters.
We do not find any significant correlation;
the largest correlation found between any foreground or beam parameter and the five-parameter $\chi^2$ was 0.12, indicating a very weak linear response and no clear direction in the foreground and beam parameter space in which the parameter comparison $\chi^2$ can be lowered.
 
We find furthermore that the foreground and beam
posteriors are not driven significantly from their priors in any chain, as one would expect if any of
these components were a poor description. The largest deviation of a parameter's posterior mean from the prior mean was $0.13\sigma$, and most parameters showed shifts of less than $0.1\sigma$.
These tests provide more support for our  hypothesis that the somewhat low PTE between $150\times150$ and PlanckFS is not due to a systematic errors in the beam or foreground treatment.

We expand on the foreground tests by adding free parameters for the kinematic Sunyaev-Zel'dovich (kSZ) effect---the S13 foreground model has a single parameter for the sum of kSZ and tSZ---and a cross-correlation between the tSZ and the cosmic infrared background (CIB) based on \cite{george15}, since these components were also included in the determination of the PlanckFS parameters.
This change has a negligible impact on the parameter posteriors for $150 \times 150$.
This result is expected: the motivation for the simplified foreground parameterization in S13 was
that these extra foreground components are expected to have a similar power-spectrum shape to tSZ
in the multipole range examined in S13 and here
and are thus only distinguishable via their frequency dependence.
Thus, the differences between \planck{} and SPT do not appear to be related to
foreground components included in the \planck{} analysis but not in S13.

\subsection{$\tau$ Prior}
Potential systematic errors may also creep in through the optical depth measurement, given the proven challenges in recovering the reionization peak from the midst of the Galactic foregrounds.
We test whether the $\tau$ prior is contributing to the low PTE by running $150\times 150$ chains with a low optical depth, $\tau=0.05\pm 0.02$, and a high optical depth,  $\tau=0.10\pm0.02$.
For the $\tau$ = 0.05 prior there are small shifts in both density parameters towards PlanckFS values but the shifts only change the PTE from \PTEPlanckfsspt to 0.058. For the $\tau$ = 0.1 prior we see the opposite effect and calculate a PTE of 0.015.
The small improvement in the PTE as $\tau$ is lowered argues against the  idea that the parameter differences between PlanckFS and $150\times 150$ are significantly connected to the $\tau$ prior.

\subsection{Gravitational Lensing}
Two of the most discrepant parameters between $150\times 150$ and PlanckFS are $\Omega_mh^2$ and $A_se^{-2\tau}$. Since these parameters both impact the lensing amplitude we test the hypothesis that the parameter differences are lensing-related. The impact of lensing on parameter estimates is often studied by marginalizing over an artifical lensing-power scaling parameter, $A_{\rm L}$. Here we follow \citet{planck16-51} and instead fix the amount of lensing power. With this choice we avoid some difficulties of interpreting parameter constraints after marginalization over $A_L$. With marginalization over $A_L$, one projects out lensing information. That is useful, but interpretation is complicated by the fact that one also removes any sensitivity to parameter variations that produce effects that can be mimicked by lensing. In contrast, by fixing the lensing potential we can remove the contribution of lensing variation to our parameter constraints, while keeping the contribution of any non-lensing responses of the power spectrum to parameter variations.

Specifically, we fix the $150\times150$ lensing potential to its best-fit \LCDM\ value. In practice we modify our model of the bandpowers (Eq. \ref{eqn:model}) so that
\begin{equation}
\begin{split}
\label{fixed_lensing_model}
M^{i\times j}_b = Y^iY^jW_{b\ell}^{i\times j}a_{\ell}\big(D^{th,UL}_{\ell}(\theta) +D^{Lens}_{\ell}(\theta_*)\\+ F_{\ell}^{i\times j}\big)
\end{split}
\end{equation}
with
\begin{equation}
    D^{Lens}_{\ell}(\theta_*)= D^{th}_{\ell}(\theta_*)-D^{th,UL}_{\ell}(\theta_*)
\end{equation}

\noindent where UL signifies an unlensed spectrum and $\theta_*$ represents cosmological parameters fixed to the $150\times150$ \LCDM\ best-fit values.

The first thing we note from the results of this test, shown in Figure \ref{fig:test_1d_params}, is that there is still some preference in the SPT data for lower matter density even with the lensing information removed, albeit a weaker preference. The increase in the size of the error bar, in the case of the fixed lensing potential, indicates that the lensing variations induced by matter density variation are important for constraints on the matter density. Adding in the lensing information strengthens the preference for low matter density.

The second thing we note is that the PTE for comparison with PlanckFS
only improves to 0.045 with lensing fixed. We attribute this to the small shift in $A_se^{-2\tau}$ that also occurs when we fix the lensing potential. Thus, although lensing has an impact, removing the impact of lensing on the SPT parameter estimates does not significantly improve agreement with the PlanckFS parameter estimates.

Finally, we note that in \citet{planck16-51} the \planck\ collaboration performed a similar test with the \planck\ data. Fixing the lensing potential lowers the matter density for \planck\ and increases it for SPT, bringing the preferred matter density values for these two datasets closer together. However, the shifts
are relatively small when compared to the full SPT and \planck\ parameter differences.

\section{Conclusions}
The \planck\ CMB temperature data at moderate angular scales ($650 \leq \ell \leq 2000$) prefer a \LCDM model that is in mild tension with some other cosmological probes.
In this paper, we have used measurements from the SPT as an independent check of the \planck\ data at these angular scales.
This check was performed by comparing \LCDM{} parameter estimates using observations of the CMB temperature anisotropies from the \planck\ satellite and SPT.
When comparing parameter constraints from the full multipole range of SPT data to parameter constraints
from \planck\ full-sky data, we found mild tension between the two, with a PTE of \PTEPlanckfsspt.
We have attempted to distinguish between three possibilities for the observed parameter differences:
slightly unusual statistical fluctuations, unaccounted-for systematic error, or a breakdown of \LCDM.
To this end, we compared parameter estimates restricted to measurements of the same modes on the sky then relaxed the range of angular scales and sky coverage.

We have arrived at three primary conclusions:
\begin{enumerate}
  \item{When \planck\ and SPT are restricted to measure the same modes on the sky (specifically, the SPT-SZ patch between $650 \leq \ell \leq 2000$), the resulting cosmological parameters are fully consistent.  This stringent test provides strong evidence against systematic contamination in either experiment at these angular scales and on this patch of sky.}
  \item{The observed tension between \planck\ on the full sky (PlanckFS) and SPT arise from both sky area (that is, the SPT-SZ patch vs.~the full sky at $650 \leq \ell \leq 2000$) and from data above $\ell>2000$ measured by SPT; however, both of these differences must be included for the difference between \planck\ and SPT parameter estimates to approach the 2$\sigma$ level.}
  \item{The high-$\ell$ SPT data (between $2000 \leq \ell \leq 3000$) drive shifts away from PlanckFS in the two density parameters $\Omega_mh^2$ and $\Omega_bh^2$---and therefore in $H_0$.  While these shifts are intriguing in the context of broader discussions of the value of $H_0$, when considered alone they are nevertheless consistent with expectations given the null hypothesis that the \LCDM model is correct and the statistical models of both data sets are accurate.}
\end{enumerate}

We arrived at these conclusions from the following set of tests and calculations.
We first quantified the tension between the best-fit \LCDM models for \planck\ and SPT and found a PTE of \PTEPlanckfssptPcnt.
We tested for systematic errors in one or both experiments by restricting \planck\ and SPT data to nearly the exact same modes on the sky. To this end, we restricted the \planck\ data to the SPT-SZ sky patch and limited each data set to the multipole range of $650\leq\ell\leq2000$.
Using the measured bandpowers and simulations described in H17 to create parameter difference covariances, we calculated $\chi^2$ values and PTEs for parameter differences between different spectra.
We found PTEs of \PTEsptcrossmin and \PTEsptplanckmin for $150\times 143$ and $143\times143$ respectively, when compared with $150\times 150$ at $\ell_{\rm max}=2000$.
This is an extremely precise test of the consistency between the two measurements, as nearly all sample variance is eliminated from the comparison.
We quantified the increased precision of this test by calculating the reduction in the volume of the 68\% confidence region for the expected distribution of parameter differences between different data comparisons;
using this metric, the $150\times 143$ vs.~$150\times150$ comparison is over 300 times more stringent than for the $150\times 150$ vs.~PlanckFS comparison.
These powerful tests would have magnified any evidence for systematic errors in either experiment;
instead, their results strongly disfavor  the presence of significant systematic errors in either the SPT or \planck{} data sets in the modes measured well by both experiments.

Next, we found that the tension between PlanckFS and SPT comes from two parts of the data.
The first is differences between the SPT-SZ patch at intermediate scales ($650 \leq \ell \leq 2000$) and the whole sky over the full range of angular scales measured by \planck\ ($\ell \leq 2000$);
this can be seen from Table~\ref{table:param PTE}, Figures \ref{fig:sim_chi_dist} and \ref{fig:dt}, and the text of Section~\ref{sec:sky_coverage}.
The second is the inclusion of high-$\ell$ data in the SPT-SZ patch ($2000 \leq \ell \leq 3000$);
this can be seen from Table~\ref{table:param SPT patch PTE}, where the PTE between PlanckFS and $150\times150$ drops below 5\% only with the inclusion of data up to $\ell_{\rm{max}}=3000$.
The tension between PlanckFS and SPT can be alleviated by removing either of these parts of the data.
By restricting \planck\ to the SPT-SZ patch, all comparisons are consistent (Table~\ref{table:param PTE});
alternately, removing the high-$\ell$ SPT data increases the PTE to \PTEPlanckfssptmin (Table~\ref{table:param SPT patch PTE}, case PlanckFS vs.~$150\times150$ with $\ell_{\rm{max}}=2000$).
Another way to say this is starting from the consistent in-patch comparison (SPT compared to \planck\ in-patch with $\ell_{\rm{max}}=2000$), if we relax either the sky coverage (PlanckFS vs.~$150\times150$ with $\ell_{\rm{max}}=2000$ $\rightarrow$ $\rm{PTE}= \PTEPlanckfssptmin $ ) or we relax the $\ell_{\rm{max}}$ range ($150\times150$ vs.~$150\times143$ $\ell_{\rm{max}}=3000$ and $150\times150$ vs.~$143\times143$ with $\ell_{\rm{max}}=3000$ $\rightarrow$ PTE = \PTEsptcross, \PTEsptplanck respectively), the datasets remain consistent; it is only when we relax \textit{both} the sky coverage and the $\ell$ range that the PTE drops below 0.05.

Third, we related certain of the parameter differences noted above to specific features in the bandpowers.
The in-patch data bandpowers have a tilt relative to PlanckFS; this can be seen from H17 and Figures~\ref{fig:best_fit_models} and \ref{fig:bandpower_model_ratio}.
This tilt is seen by both \planck\ and SPT data, and is thus unlikely to arise from systematics in this range of angular scales in either experiment.
We find that this tilt couples to the \LCDM parameters $A_se^{-2\tau}$, and $n_s$ (see Section~\ref{sec:bandpower_ratios} and Figure~\ref{fig:test_1d_params}).
The tilt does not couple to the density parameters and $H_0$;
this is confirmed by the fact that when the tilt is artifically removed, $H_0$ remains high relative to PlanckFS (see Section~\ref{sec:bandpower_ratios} and Figure~\ref{fig:test_1d_params}).

Finally, we performed an additional set of tests designed to investigate whether specific potential sources
of systematic error could be responsible for any of the measured parameter differences. We investigated
the effects of the SPT instrument beam, the treatment of foregrounds in SPT data, and influence of
assumptions about the optical depth to reionization and the amplitude of gravitational lensing in the
analysis. We found no evidence of any coupling of these effects to the measured parameter differences.

We conclude that, at most, our tests reveal weak evidence for a breakdown of \LCDM or systematic errors influencing either the \planck\ data outside the SPT-SZ survey area or the SPT data at $\ell >2000$.
Instead, the tension between SPT and \planck\, under \LCDM can plausibly be explained by two
individually insignificant statistical fluctuations---one between the SPT-SZ survey area and the full sky, the other in the high $\ell$ data better constrained by SPT.

Whether this explanation is correct will ultimately be determined most directly by additional observations of the CMB temperature anisotropies at $\ell > 2000$, both within and beyond the SPT-SZ patch.
Additionally, measurements of the $EE$ and $TE$ CMB polarization power spectra---from, e.g., Advanced ACTPol \citep{henderson16}, SPT-3G \citep{benson14}, and CMB-S4 \citep{abazajian16}---will provide yet more stringent tests of our standard cosmological model.

\acknowledgements{The South Pole Telescope is supported by the National Science Foundation through grant PLR-1248097.  Partial support is also provided by the NSF Physics
Frontier Center grant PHY-1125897 to the Kavli Institute of Cosmological Physics at the University of Chicago, the Kavli Foundation and the Gordon and
Betty Moore Foundation grant GBMF 947. B. Benson was supported by the Fermi Research Alliance, LLC under Contract No. DE-AC02-07CH11359 with the U.S. Department of Energy, Office of Science, Office of High Energy Physics. C. Reichardt acknowledges support from an Australian Research Council Future Fellowship (FT150100074). The McGill group acknowledges funding from the National Sciences and Engineering Research Council of Canada, Canada Research Chairs program, and the Canadian Institute for Advanced Research.
Work at Argonne National Laboratory was supported under U.S. Department of Energy contract DE-AC02-06CH11357. This work used resources made available on the Jupiter cluster, a joint data-intensive computing project between the High Energy Physics Division and the Computing, Environment, and Life Sciences (CELS) Directorate at Argonne National Laboratory.
We thank G. Addison for pointing out the importance of the aberration correction.}

\begin{appendix}
\section{On the Calculation of Calibration and Beam Uncertainty}

In this appendix, we discuss  why we use a different method for handling beam and calibration uncertainty than what was done in S13. We discovered that the method used in S13 leads to a bias lowering the estimates for $A_se^{-2\tau}$ and $n_s$. We explain why the method used in S13 leads to a bias and present an unbiased procedure to include beam and calibration uncertainty in the likelihood.

In S13, the calibration and beam uncertainty of the bandpowers was added  into the full covariance ($\vec{\Sigma}$) in a data ($\vec{D}$) dependent way as

\begin{align}
    \Sigma_{bb'} &= \Sigma_{bb'}^S+\Sigma_{bb'}^N+D_{b}D_{b'}W_{b\ell}W_{b'\ell'}\rho^{B}_{\ell\ell'} + D_{b}D_{b'}\sigma_Y^2
\end{align}

\noindent where S, and N signify the sample and noise covariances, $W_{b\ell}$ are the window functions, the beam correlation term, $\rho^{B}_{\ell\ell'}$, is formed in a manner similar to Eq.~\ref{eqn:beam_corr}, and $\sigma_Y$ is the calibration uncertainty.

Using the data in the calculation of the covariance instead of a fiducial model introduces a bias in the likelihood. Elements of $\vec{D}$ with smaller values will have less beam and calibration uncertainty than larger elements of $\vec{D}$. When fitting a model to $\vec{D}$, the preference will be to fit the smaller valued elements of $\vec{D}$ better than larger elements. The calibration uncertainty bias has a greater effect at low-$\ell$ and the beam uncertainty bias has a greater effect at high-$\ell$, explaining the connection to the parameters $A_se^{-2\tau}$ and $n_s$. If the beam and calibration uncertainty are added into the covariance in a model-dependent way, the errorbars on all elements of $\vec{D}$ are adjusted with the model.
The resulting fits are unbiased, since a model that favors fitting the smallest elements of $\vec{D}$  produces smaller errorbars for all elements of $\vec{D}$ and gives a worse overall fit.

This bias can best be understood in the context of a simple example. Since the bias works in the same manner for both beam and calibration uncertainty, we focus on just the latter. We assume a set of data ($\vec{d}$) from a Gaussian distribution  with covariance $\vec{C}$. Our model will take the form $Y\vec{m}(\theta)$, where we explicitly include the calibration $Y$, and $\vec{m}(\theta)$  depends on the remaining parameters in our model ($\theta$).

With a Gaussian prior on $Y$ of $N(1,\sigma_Y)$, the probability for the model parameters, $\theta$ and $Y$ given $\vec{d}$, can be written as
\begin{align}
    P(\theta,Y|\vec{d})=\frac{P(\vec{d}|\theta,Y)P(Y)}{P(\vec{d})} \propto e^{-\frac{1}{2}(\vec{d}- Y\vec{m})^{\dagger}\vec{C}^{-1}(\vec{d}-Y \vec{m}) }e^{-\frac{(Y-1)^2}{2\sigma_{Y}^2}}.
\end{align}
If we analytically marginalize over $Y$ we get
\begin{align}
    P(\theta|\vec{d}) =\int_{-\infty}^{\infty}dYP(\theta,Y|\vec{d})\propto \int_{-\infty}^{\infty}dY
     e^{-\frac{1}{2}(\vec{d}-Y \vec{m})^{\dagger}\vec{C}^{-1}(\vec{d}-Y\vec{m})}e^{-\frac{(Y-1)^2}{2\sigma_{Y}^2}}
\end{align}
After integrating over $Y$ and working through some algebra, we have
\begin{align}
    P(\theta|\vec{d}) \propto exp\Bigg(-\frac{1}{2}\Bigg[ \vec{d}^{\dagger}\vec{C}^{-1}\vec{d} + \frac{1}{\sigma_{Y}^2} - \frac{\Big(\vec{m}^{\dagger}\vec{C}^{-1}\vec{d}+\frac{1}{\sigma_{Y}^2}\Big)^2}{\vec{m}^{\dagger}\vec{C}^{-1}\vec{m} + \frac{1}{\sigma_{Y}^2}}\Bigg]\Bigg)
\end{align}
At this point, we only care about the term in brackets in Eq. A4 and wish to recast it in the form of $\chi^2 = (\vec{d}-\vec{x})^{\dagger}\vec{C}'^{-1}(\vec{d}-\vec{x})$. Rearranging terms gives
\begin{align}
    \chi^2 = \vec{d}^{\dagger}\Bigg[\vec{C}^{-1} - \frac{\vec{C}^{-1}\vec{m}\vec{m}^{\dagger}\vec{C}^{-1}}{\vec{m}^{\dagger}\vec{C}^{-1}\vec{m} + \frac{1}{\sigma_{Y}^2}}\Bigg]\vec{d}- 2\vec{m}^{\dagger}\Bigg[\frac{\vec{C}^{-1}}{\sigma_{Y}^2\vec{m}^{\dagger}\vec{C}^{-1}\vec{m}+1}\Bigg]\vec{d} - \frac{1}{\sigma_{Y}^4\vec{m}^{\dagger}\vec{C}^{-1}\vec{m}+\sigma_{Y}^2}+\frac{1}{\sigma_{Y}^2}
\end{align}
The term quadratic in $\vec{d}$ gives us $\vec{C}'^{-1}$. Using the Sherman-Morrison-Woodbury formula tells us
\begin{align}
 \vec{C}'=\vec{C}+\sigma_{Y}^2\vec{m}\vec{m}^{\dagger}.
\end{align}

The correct way to include the calibration uncertainty into the covariance is by multiplying the uncertainty by the outer product of the model. If one were to run MCMC with the calibration marginalized over, then Eq. A6 would need to be updated with $\vec{m}$ at each step.

To see how using the data ($\vec{d}$) in Eq. A6 instead of $\vec{m}$ would lead to a bias, lets consider just two data points, one scattered high and one scattered low from the truth with equal variance.
If we wished to model the two data points as a constant line, the best-fit in terms of minimizing $\chi^2$ would be a horizontal line, $m_i=m\mu_i$ where $\mu_i=1$ $\forall$ $i$ and $m=\frac{1}{2}(d_1+d_2)$, that splits the difference between the two points. If we instead make the calibration uncertainty data dependent,
the covariance becomes
\begin{align}
\vec{C}''=\vec{C}+\sigma_{Y}^2\vec{d}\vec{d}^{\dagger}
\end{align}
and minimizing $\chi^2$ tells us
\begin{align}
m = \frac{\sum_{i,j}(\vec{C}''^{-1})_{ij}d_j}{\sum_{i,j}(\vec{C}''^{-1})_{ij}}.
\end{align}

\noindent The best-fit line will be shifted towards the lower valued data point since it receives a greater weighting from $\vec{C}''^{-1}$. The magnitude of the shift depends on the magnitude of the calibration uncertainty.

To avoid the bias discussed in this section, we multiply our model by a calibration parameter and include a prior in our likelihood. Beam uncertainty is either incorparated in a model dependent way similar to Eq. A6 or as parameterized templates.

\end{appendix}
\bibliography{spt.bib}

\end{document}

%% file: spt_authorlist_v2.tex
\def\Davis{1}
\def\KICPChicago{2}
\def\AAUChicago{3}
\def\KIPAC{4}
\def\Stanford{5}
\def\FNAL{6}
\def\ArgonneHEP{7}
\def\PhysicsUChicago{8}
\def\EFIChicago{9}
\def\SLAC{10}
\def\McGill{11}
\def\Caltech{12}
\def\Berkeley{13}
\def\Cifar{14}
\def\Colorado{15}
\def\ESO{16}
\def\Colphys{17}
\def\Illast{18}
\def\Illphys{19}
\def\UChicago{20}
\def\LBNL{21}
\def\Arizona{22}
\def\Michigan{23}
\def\Munich{24}
\def\ExcellenceCluster{25}
\def\MPE{26}
\def\Dunlap{27}
\def\Minnesota{28}
\def\Melbourne{29}
\def\CaseWestern{30}
\def\ArtInstChicago{31}
\def\JPL{32}
\def\CfA{33}

\def\UToronto{34}

\author{
  K.~Aylor\altaffilmark{\Davis},
  Z.~Hou\altaffilmark{\KICPChicago,\AAUChicago},
  L.~Knox\altaffilmark{\Davis},
  K.~T.~Story\altaffilmark{\KIPAC,\Stanford},
  B.~A.~Benson\altaffilmark{\FNAL,\KICPChicago,\AAUChicago},
  L.~E.~Bleem\altaffilmark{\ArgonneHEP,\KICPChicago},
  J.~E.~Carlstrom\altaffilmark{\KICPChicago,\ArgonneHEP,\PhysicsUChicago,\AAUChicago,\EFIChicago},
  C.~L.~Chang\altaffilmark{\ArgonneHEP,\KICPChicago,\AAUChicago},
  H-M.~Cho\altaffilmark{\SLAC},
  R.~Chown\altaffilmark{\McGill},
  T.~M.~Crawford\altaffilmark{\KICPChicago,\AAUChicago},
  A.~T.~Crites\altaffilmark{\KICPChicago,\AAUChicago,\Caltech},
  T.~de~Haan\altaffilmark{\McGill,\Berkeley},
  M.~A.~Dobbs\altaffilmark{\McGill,\Cifar},
  W.~B.~Everett\altaffilmark{\Colorado},
  E.~M.~George\altaffilmark{\Berkeley,\ESO},
  N.~W.~Halverson\altaffilmark{\Colorado,\Colphys},
  N.~L.~Harrington\altaffilmark{\Berkeley},
  G.~P.~Holder\altaffilmark{\McGill,\Cifar,\Illast,\Illphys},
  W.~L.~Holzapfel\altaffilmark{\Berkeley},
  J.~D.~Hrubes\altaffilmark{\UChicago},
  R.~Keisler\altaffilmark{\KICPChicago,\PhysicsUChicago,\KIPAC},
  A.~T.~Lee\altaffilmark{\Berkeley,\LBNL},
  E.~M.~Leitch\altaffilmark{\KICPChicago,\AAUChicago},
  D.~Luong-Van\altaffilmark{\UChicago},
  D.~P.~Marrone\altaffilmark{\Arizona},
  J.~J.~McMahon\altaffilmark{\Michigan},
  S.~S.~Meyer\altaffilmark{\KICPChicago,\AAUChicago,\EFIChicago,\PhysicsUChicago},
  M.~Millea\altaffilmark{\Davis},
  L.~M.~Mocanu\altaffilmark{\KICPChicago,\AAUChicago},
  J.~J.~Mohr\altaffilmark{\Munich,\ExcellenceCluster,\MPE},
  T.~Natoli\altaffilmark{\Dunlap},
  Y.~Omori\altaffilmark{\McGill},
  S.~Padin\altaffilmark{\KICPChicago,\AAUChicago},
  C.~Pryke\altaffilmark{\Minnesota},
  C.~L.~Reichardt\altaffilmark{\Berkeley,\Melbourne},
  J.~E.~Ruhl\altaffilmark{\CaseWestern},
  J.~T.~Sayre\altaffilmark{\CaseWestern,\Colorado},
  K.~K.~Schaffer\altaffilmark{\KICPChicago,\EFIChicago,\ArtInstChicago},
  E.~Shirokoff\altaffilmark{\Berkeley,\KICPChicago,\AAUChicago}, 
  Z.~Staniszewski\altaffilmark{\CaseWestern,\JPL},
  A.~A.~Stark\altaffilmark{\CfA},
  K.~Vanderlinde\altaffilmark{\Dunlap,\UToronto},
  J.~D.~Vieira\altaffilmark{\Illast,\Illphys}, and
  R.~Williamson\altaffilmark{\KICPChicago,\AAUChicago}
  }

\altaffiltext{\Davis}{Department of Physics, University of California, Davis, CA, USA 95616}
\altaffiltext{\KICPChicago}{Kavli Institute for Cosmological Physics, University of Chicago, Chicago, IL, USA 60637}
\altaffiltext{\AAUChicago}{Department of Astronomy and Astrophysics, University of Chicago, Chicago, IL, USA 60637}
\altaffiltext{\KIPAC}{Kavli Institute for Particle Astrophysics and Cosmology, Stanford University, 452 Lomita Mall, Stanford, CA 94305}
\altaffiltext{\Stanford}{Dept. of Physics, Stanford University, 382 Via Pueblo Mall, Stanford, CA 94305}
\altaffiltext{\FNAL}{Fermi National Accelerator Laboratory, MS209, P.O. Box 500, Batavia, IL 60510}
\altaffiltext{\ArgonneHEP}{High Energy Physics Division, Argonne National Laboratory, Argonne, IL, USA 60439}
\altaffiltext{\PhysicsUChicago}{Department of Physics, University of Chicago, Chicago, IL, USA 60637}
\altaffiltext{\EFIChicago}{Enrico Fermi Institute, University of Chicago, Chicago, IL, USA 60637}
\altaffiltext{\SLAC}{SLAC National Accelerator Laboratory, 2575 Sand Hill Road, Menlo Park, CA 94025}
\altaffiltext{\McGill}{Department of Physics and McGill Space Institute, McGill University, Montreal, Quebec H3A 2T8, Canada}
\altaffiltext{\Caltech}{California Institute of Technology, Pasadena, CA, USA 91125}
\altaffiltext{\Berkeley}{Department of Physics, University of California, Berkeley, CA, USA 94720}
\altaffiltext{\Cifar}{Canadian Institute for Advanced Research, CIFAR Program in Cosmology and Gravity, Toronto, ON, M5G 1Z8, Canada}
\altaffiltext{\Colorado}{Center for Astrophysics and Space Astronomy, Department of Astrophysical and Planetary Sciences, University of Colorado, Boulder, CO, 80309}
\altaffiltext{\ESO}{European Southern Observatory, Karl-Schwarzschild-Stra{\ss}e 2, 85748 Garching, Germany}
\altaffiltext{\Colphys}{Department of Physics, University of Colorado, Boulder, CO, 80309}
\altaffiltext{\Illast}{Astronomy Department, University of Illinois at Urbana-Champaign, 1002 W. Green Street, Urbana, IL 61801, USA}
\altaffiltext{\Illphys}{Department of Physics, University of Illinois Urbana-Champaign, 1110 W. Green Street, Urbana, IL 61801, USA}
\altaffiltext{\UChicago}{University of Chicago, Chicago, IL, USA 60637}

\altaffiltext{\LBNL}{Physics Division, Lawrence Berkeley National Laboratory, Berkeley, CA, USA 94720}
\altaffiltext{\Arizona}{Steward Observatory, University of Arizona, 933 North Cherry Avenue, Tucson, AZ 85721}
\altaffiltext{\Michigan}{Department of Physics, University of Michigan, Ann  Arbor, MI, USA 48109}
\altaffiltext{\Munich}{Faculty of Physics, Ludwig-Maximilians-Universit\"{a}t, 81679 M\"{u}nchen, Germany}
\altaffiltext{\ExcellenceCluster}{Excellence Cluster Universe, 85748 Garching, Germany}
\altaffiltext{\MPE}{Max-Planck-Institut f\"{u}r extraterrestrische Physik, 85748 Garching, Germany}
\altaffiltext{\Dunlap}{Dunlap Institute for Astronomy \& Astrophysics, University of Toronto, 50 St George St, Toronto, ON, M5S 3H4, Canada}
\altaffiltext{\Minnesota}{Department of Physics, University of Minnesota, Minneapolis, MN, USA 55455}
\altaffiltext{\Melbourne}{School of Physics, University of Melbourne, Parkville, VIC 3010, Australia}
\altaffiltext{\CaseWestern}{Physics Department, Center for Education and Research in Cosmology and Astrophysics, Case Western Reserve University,Cleveland, OH, USA 44106}
\altaffiltext{\ArtInstChicago}{Liberal Arts Department, School of the Art Institute of Chicago, Chicago, IL, USA 60603}
\altaffiltext{\JPL}{Jet Propulsion Laboratory, California Institute of Technology, Pasadena, CA 91109, USA}
\altaffiltext{\CfA}{Harvard-Smithsonian Center for Astrophysics, Cambridge, MA, USA 02138}
\altaffiltext{\UToronto}{Department of Astronomy \& Astrophysics, University of Toronto, 50 St George St, Toronto, ON, M5S 3H4, Canada}